\documentclass[conference]{IEEEtran}

\usepackage[utf8]{inputenc} 
\usepackage[T1]{fontenc}
\usepackage{url}
\usepackage{ifthen}
\usepackage{cite}
\usepackage[cmex10]{amsmath} 

\newtheorem{theorem}{Theorem}

\newtheorem{cor}{Corollary}
\newtheorem{lem}{Lemma}

\ifCLASSINFOpdf
   \usepackage[pdftex]{graphicx}
  \else
   \usepackage[dvipdfmx]{graphicx}
  \fi
\usepackage{color}
\usepackage{amssymb}
\usepackage{amsfonts}
\usepackage{bm}

\begin{document}
%

\title{Connectivity of Ad Hoc Wireless Networks with Node Faults}

\author{\IEEEauthorblockN{Satoshi Takabe and Tadashi Wadayama}
\IEEEauthorblockA{Nagoya Institute of Technology,\\
                    Gokiso-cho, Showa-ku, Nagoya,
                    Aichi, Japan\\
                    Email: \{s\_takabe, wadayama\}@nitech.ac.jp}
}
\maketitle

\begin{abstract}
Connectivity of wireless sensor networks (WSNs) is a fundamental global property expected to be maintained 
even though some sensor nodes are at fault.
In this paper, we investigate {the} connectivity of random geometric graphs (RGGs) in the node fault model
 as an abstract model of ad hoc WSNs with unreliable nodes.
 In the model, each node is assumed to be stochastically at fault, i.e., removed from a graph.
As a measure of reliability, the network breakdown probability is then defined as 
the average probability that a resulting survival graph is disconnected over RGGs.
We examine RGGs with general connection functions 
as an extension of a conventional RGG model and 
provide two mathematical analyses:
the asymptotic analysis for infinite RGGs that reveals the phase transition thresholds of connectivity,
 and the non-asymptotic analysis for finite RGGs that provides a useful approximation formula.
Those analyses are supported by numerical simulations in the Rayleigh SISO model
 reflecting a practical wireless channel.
\end{abstract}


\IEEEpeerreviewmaketitle

\section{Introduction}

Recent progress on IoT technologies has promoted the extensive studies on the wireless sensor networks (WSNs).
The WSN contains a large number of sensor nodes with several sensors and a small transceiver to communicate to other sensor nodes.
It is highly desirable for a WSN to maintain its {\em connectivity}~\cite{Zhu} because it ensures successful node-to-node communications 
over the WSN. In some cases, a sensor node is deployed in a harsh environment which is not suitable for an electric device,
and a sensor node has only restricted energy {resources} such as small batteries or {solar cells~\cite{Sha}}. 
A malfunction or battery shortage occurred on a sensor node is not a rare event for such a WSN.
 In other words, we should regard nodes in a WSN as {\em unreliable nodes} when we assess the robustness
 and immunity of the network against malfunctions or battery shortage.
  In this context, it would be natural to ask the relationship between local model parameters such as the transmit power 
of each node and a global property of the WSN such as connectivity. It is expected that increasing the transmit power 
leads to a more robust network against the node faults,
{ and to} fast energy consumption. The trade-offs between them 
is worth exploring because it helps us to design 
an energy efficient WSN~\cite{Haen} that is immune to the node faults.

In many cases, {\em ad hoc} WSNs are created by a random deployment of sensor nodes in a target area.
To study qualitative and qualitative natures of ad hoc WSNs, 
random geometric graphs (RGGs) are commonly used as a standard abstract model~\cite{Gil}.
A RGG is defined by a point process corresponding to a random node deployment
and by a random edge assignment according to a stochastic rule.
There are a number of RGG models with different rules for random edge assignments.
These models are basically characterized 
by a \emph{connection function}, i.e., the probability of edge assignment
as a function of the distance between two nodes.
For example, if we simply assume that each node is 
connected to neighboring nodes within a fixed distance,
the corresponding RGG model is sometimes called the \emph{hard-disk model}.
The connectivity problem of the hard-disk model is first introduced by 
Krishnamachari et al.~\cite{Kris}.
{They numerically studied the existence of the phase transition, which is mathematically proved 
in random graph theory~\cite{Det1}.}
The phase transition of connectivity has been studied in more practical RGG models~\cite{Bet,Hek,Fabbri}.
Recently, Mao and Anderson showed the phase transition phenomena of connectivity
 in {infinite} RGGs with general connection functions~\cite{Mao}
 while a general approximation theory for finite {RGGs was also} proposed in~\cite{Dett}.
 
A random graph with unreliable nodes is called a {\em node fault model} in the present paper.
The node fault model has been studied mainly for graphs which is not 
embedded in a metric space.
The model was originally studied in~\cite{The} as {\em imperfect networks}.
Nozaki et al. obtained an upper bound of the network breakdown probability
 averaged over regular random graphs~\cite{nnw}.
Stimulated by this work, the authors proposed an approximation formula for an arbitrary random graph
 characterized by the degree distribution~\cite{tnw}.
For RGGs, on the other hand, Wan and Yi studied the hard-disk model~\cite{Peng}.
They obtained the critical transmission range, i.e., the phase transition threshold, of connectivity
using geometric and probabilistic evaluations.

The main goal of this paper is to clarify 
the relationship between local parameters of RGG models such as transmit power
and the connectivity of RGGs with a general connection function and unreliable nodes.
As described above, the RGG model is characterized by a connection function that reflects
the environment in which WSNs are deployed, i.e., fading, shadowing, 
and the degree of scattering. This paper will provide a general framework to examine 
a connectivity issue in WSNs, which is useful for WSN design.
In this paper, we assume a simple node fault model such that a node fault takes place independently 
with constant probability $\epsilon$. The probability that a {\em survival graph} is connected
 after probabilistic node breakdowns is called the {\em network breakdown probability}.

We first study the asymptotic situation in which the number of nodes goes to infinity because
such a problem setting illuminates the fundamental nature of the system.
Recently, the authors obtained asymptotic upper and lower bounds 
of the network breakdown probability for random graphs and RGGs \cite{tw}.
 By combining this result and the results in~\cite{Mao}, we will show the existence of the phase transition
 regarding the network breakdown probability and derive its threshold.
 
We then focus on the non-asymptotic case where the number of nodes is finite.
The setup is closer to practical WSNs.
A simple approximation formula of the network breakdown probability will be derived 
based on the result {by} Dettmann and Georgiou~\cite{Dett}. The approximation formula 
is a useful tool to assess the robustness of the network with a given set of parameters.
Several numerical simulations were carried out for evaluating the tightness of
the approximation formula under the assumption of the Rayleigh SISO model.

\section{Preliminaries}

\subsection{Notations}
An event $\mathcal{A}_n$ depending on an integer $n$ is said to occur 
asymptotically almost surely (a.a.s.) if its probability converges to one as $n\rightarrow \infty$.
We define big-O notations for a function $f$ as follows: 
$f(x)=O(g(x))$ iff $\mathrm{limsup}_{x\rightarrow\infty}|f(x)|/g(x) <\infty$ holds,
$f(x)=\Theta(g(n))$ iff there exist positive constants $c_1$ and $c_2$ such that 
$c_1g(x)\le f(x)\le c_2g(x)$ holds for sufficient large $x$, and 
$f(x)=o(g(n))$ iff $f(x)/g(x)\rightarrow 0$ ($n\rightarrow\infty$) holds.

\subsection{Random Geometric Graphs}\label{rgg}

The RGG is a random graph model defined in a metric space.
In this paper, we specifically consider the two-dimensional Euclidean space $\mathbb{R}^2$.
Each node in a RGG is randomly deployed in {a bounded closed domain  
$S \subset \mathbb{R}^2$}.
As a point process, we use the uniform point process $\mathcal{X}_n$ with $n$ nodes in which each node is 
independently and uniformly deployed in $S$.
The simplest RGG model $\mathcal{G}(\mathcal{X}_n,r)$ referred to
the hard-disk model is then defined
 by setting edges to a pair of nodes whose distance is at most $r(>0)$~\cite{Peng}.

We in this paper deal with the {\em general connection model} $\mathcal{G}(\mathcal{X}_n,g_{r_n})$~\cite{Mee}
 that includes the hard-disk model and other practical models.
In the general connection model, each edge is assigned to a pair of nodes with probability $g_{r_n}(r)$
where $r$ is the distance between those nodes and
the parameter $r_n$ is related to the appropriate length scale regarding 
the phase transition (see Theorem \ref{thm_Mao}) 
generally depending on the number of nodes.

As described in~\cite{Mao}, we assume that the connection function $g_{r_n}$ is rotationally invariant and rescaled as follows:
\begin{equation}
g_{r_n}(r)=g\left(\frac{r}{r_n}\right), \label{eq_cond1}
\end{equation}
where a function $g:[0,\infty)\rightarrow [0,1]$ characterizes an environment in which WSNs are located.
We further assume that the function $g$ satisfies monotonicity and integral boundedness which are respectively given as
\begin{align}
&g(x)\ge g(y) \quad \mbox{if } x\le y,\label{eq_cond2}\\
&0< C\triangleq \int_{\mathbb{R}^2}d\bm{x}g(\|\bm{x}\|) <\infty,
\label{eq_rgg_a}
\end{align}
where $\|\bm{x}\|$ represents the Euclidean norm of $\bm{x}$.
The phase transition threshold of connectivity depends on $C$ defined in~(\ref{eq_rgg_a})
 as shown in the next section.
In addition, we assume that the function $g$ decreases sufficiently rapidly, i.e.,
\begin{equation}
g(x)=o(1/(x^2\ln^2 x)).\label{eq_cond3}
\end{equation}
The hard-disk model $\mathcal{G}(\mathcal{X}_n,r)$ is recovered if we choose $g(x)=\theta(1-x)$
 where $\theta(x)$ takes $1$ if $x\ge 1$ and $0$ otherwise.
The general connection functions allow us 
to handle more practical ad hoc WSN models as summarized in~\cite{Dett}.

\subsection{Node Fault Model}

We briefly describe the node fault model and define the network breakdown probability.
General and formal definitions are found in~\cite{tw}.

The \emph{node fault model with node breakdown probability} $\epsilon$ is a stochastic process
 that each node in a graph $G=(V,E)$ is independently at fault with a constant probability $\epsilon\in[0,1)$.
A subset $V_b$ of the node set $V$ is defined as a set of fault nodes.
A \emph{survival graph} of $G$ is then defined as the induced subgraph of 
the survival nodes $V\backslash V_b$.
In a WSN, a fault node corresponds to a broken sensor node that cannot relay packets.
We thus are interested in the connectivity of the survival graph.

For the node fault model 
with the node breakdown probability $\epsilon$,
the network breakdown probability $P_b(G, \epsilon)$ 
of $G$ is defined as the probability that  the survival graph is not connected.
We then define the {\em expected  network breakdown probability} 
${P}_{\Omega_n}(\epsilon)$ as the breakdown probability $P_b(G, \epsilon)$
averaged over a random graph model $\Omega_n$ with $n$ nodes.
{In this paper, we simply call ${P}_{\Omega_n}(\epsilon)$ the network breakdown probability.}
{We can expect that this average reflects the typical value of 
the network breakdown probability of any instance in a random graph model if 
the number of nodes is sufficiently large.}

\section{Asymptotic analysis}\label{sec_prob}

\subsection{Connectivity of RGGs with general connection function}\label{con_normal}
Connectivity of RGGs with the general connection function 
was investigated by Mao and Anderson~\cite{Mao}.
They studied a random connection model $\mathcal{G}(\mathcal{P}_\rho,g_{r_\rho})$
 with the Poisson point process $\mathcal{P}_\rho$ with density $\rho$.
The Poisson point process $\mathcal{P}_\rho$ in the domain $S$ 
is defined as a node random deployment following Poisson distribution with density $\rho$.
They proved the following phase transition phenomenon with respect to connectivity.
\begin{theorem}[\cite{Mao}, Thm. 9, Thm. 10]\label{thm_Mao} 
Assume that $S$ is the unit-area square.
Let
\begin{equation}
r_\rho^\ast\triangleq \sqrt{\frac{\ln \rho+b}{C\rho}},
\end{equation}
with a factor $C$ in~(\ref{eq_rgg_a}) and a constant $b$.
The functions $g_{r_n}$ and $g$ satisfy the conditions (\ref{eq_cond1})-(\ref{eq_cond3}).
Then, as $\rho\rightarrow\infty$, the probability that $\mathcal{G}(\mathcal{P}_\rho,g_{r_\rho^\ast})$ is connected
is $e^{-e^{-b}}$.
Especially, as $\rho\rightarrow\infty$, $\mathcal{G}(\mathcal{P}_\rho,g_{r_\rho^\ast})$ is a.a.s. connected iff $b\rightarrow\infty$
while $\mathcal{G}(\mathcal{P}_\rho,g_{r_\rho^\ast})$ is a.a.s. disconnected iff $b\rightarrow-\infty$.
\end{theorem}
{This theorem indicates that 
the specific scale factor $r_\rho$ characterizes the phase transition threshold
 and unveils the connection between local  and global properties of
our interests.}

\subsection{de-Poissonization}

In this paper, we focus on the uniform $n$-point process $\mathcal{X}_n$ instead of
the Poisson point process $\mathcal{P}_n$ with density $n$
 because the number of sensor nodes is to be fixed in practice.
This also simplifies the following discussions.
In RGGs with $\mathcal{P}_n$,
the number of nodes can be fluctuated while those of $\mathcal{X}_n$ is fixed to $n$.
This fact affects the connection probability of random graphs with finite nodes.
However, $\mathcal{P}_n$ is similar to $\mathcal{X}_n$ in terms of mutually independent property: 
given that $A_1,A_2,\dots,A_k$ ($k=1,2,\dots$) be an arbitrary set of
 disjoint regions in $S$, the process $\mathcal{P}_n$ deploys nodes such that 
the number of nodes in $A_1,A_2,\dots,A_k$ are mutually independent random variables with Poisson distributions with density 
$n|A_1|,n|A_2|,\dots,n|A_k|$, respectively.
It enables us to approximate RGGs with $\mathcal{X}_n$ to those with $\mathcal{P}_n$ and vice versa
 for sufficient large $n$.
This technique was first proposed by Penrose~\cite{Pen97} and is called the \emph{(de-)Poissonization technique}.

The de-Poissonization technique is applicable to Thm.~\ref{thm_Mao} and related theorems for the general connection model.
We thus immediately obtain the following theorem from Thm.~\ref{thm_Mao}.
\begin{theorem}\label{thm_con} 
Assume that $S$ be the unit-area square and
\begin{equation} \label{rn1}
r_n^\ast\triangleq \sqrt{\frac{\ln n+b}{Cn}},
\end{equation}
with a factor $C$ in~(\ref{eq_rgg_a}) and a constant $b$.
The functions $g_{r_n}$ and $g$ satisfy the conditions (\ref{eq_cond1})-(\ref{eq_cond3}).
Then, as $n\rightarrow\infty$, the probability that $\mathcal{G}(\mathcal{X}_n,g_{r_n^\ast})$ is connected
is given by $e^{-e^{-b}}$.
Especially, as $n\rightarrow\infty$, $\mathcal{G}(\mathcal{X}_n,g_{r_n^\ast})$ is a.a.s. connected iff $b\rightarrow\infty$
while $\mathcal{G}(\mathcal{X}_n,g_{r_n^\ast})$ is a.a.s. disconnected iff $b\rightarrow-\infty$.
\end{theorem}

This theorem plays a key role to derive the result on the phase transition {in} 
the node fault model to be presented.
Recall that the factor $C$ depends only on the connection function $g$
and a constant with respect to the length scale $r_n$ and the number of nodes $n$.
In other words, the difference of the connection functions in RGG models has influence
only on a constant factor with respect to the threshold.

\subsection{Connectivity in node fault model}\label{con}

We turn back to the node fault model.
In~\cite{tw}, the authors obtained asymptotic bounds generally applicable to a class of RGGs. 
We briefly review a part of results in \cite{tw} that is required for the following discussion.

In the node fault model, the cardinality of a survival graph, denoted by  $s$, 
fluctuates around its mean $(1-\epsilon) n$.
We then define a network breakdown probability $P_{\Omega_n}(\epsilon;s)$ $(0\le s\le n)$
with $s$ nodes as a conditional probability of $P_{\Omega_n}(\epsilon)$ that a survival graph with $s$ nodes is disconnected.
From the sum rule of conditional probabilities $P_{\Omega_n}(\epsilon;s)$, we have the following 
asymptotic bound of the network breakdown probability \cite{tw}.
\begin{lem}\label{lem_typ_bre}
Let $\delta_n$ be a sequence for all $n$ which satisfies $\delta_n\rightarrow 0$ and $\delta_n\sqrt{n}\rightarrow \infty$
 as $n\rightarrow\infty$~\footnote{Conditions on a sequence $\delta_n$ are satisfied if, e.g., $\delta_n=n^{-1/3}$.}.
Then, for sufficiently large $n$, the network breakdown probability satisfies 
\begin{equation}
\begin{aligned}
\left(1-\frac{1}{2n\delta_n}\right)&\min_{s\in[s^{(-)}, s^{(+)}]}P_{\Omega_n}(\epsilon;s)
\le P_{\Omega_n}(\epsilon)\\
&\quad\quad\le\frac{1}{2n\delta_n}+\max_{s\in[s^{(-)}, s^{(+)}]}P_{\Omega_n}(\epsilon;s), 
\end{aligned}
\label{eq_typ4}
\end{equation}
where $s^{(\pm)}\triangleq (\kappa\pm\delta_n)n$ and $\kappa\triangleq 1-\epsilon$.
\end{lem}

We then consider the general connection model $\mathcal{G}(\mathcal{X}_n,g_{r_n})$.
For $\mathcal{G}(\mathcal{X}_n,g_{r_n})$, the ensemble of survival graphs with $s$ nodes
is equivalent to a general connection model $\mathcal{G}(\mathcal{X}_s,g_{r_n})$ with $s$ nodes.
 This claim is based on the fact that the edge assignment process in the general connection model
  and the random node deletion process in the node fault model is exchangeable.
 Exchanging those processes defines a new process that $n-s$ nodes are uniformly removed after the process $\mathcal{X}_n$,
 which is equivalent to $\mathcal{X}_s$.
 We thus conclude the equivalence of ensembles.
It enables us to evaluate the maximum and minimum of $P_{\Omega_n}(\epsilon;s)$ in $s\in [s^{(-)},s^{(+)}]$.
We have the following theorem according to the phase transition of the network breakdown probability
when the node breakdown probability $\epsilon$ is fixed.

\begin{theorem}\label{prop_rrg1} 
Let us consider the node fault model with node breakdown probability $\epsilon\in [0,1)$.
Assume that $S$ be the unit-area square, 
\begin{equation} \label{rn2}
r_n^\ast\triangleq \sqrt{\frac{\ln n+b}{C\kappa n}},
\end{equation}
with a factor $C$ in~(\ref{eq_rgg_a}) and a constant $b$, and
functions $g_{r_n}$ and $g$ satisfy conditions (\ref{eq_cond1})-(\ref{eq_cond3}).
Then, for $\mathcal{X}_n$ over $S$ and $\Omega_n=\mathcal{G}(\mathcal{X}_n,g_{r_n^\ast})$, 
${P}_{\Omega_n}(\epsilon)=1-e^{-\kappa e^{-b}}$ as $n\rightarrow\infty$.
Especially, as $\rho\rightarrow\infty$, the survival graph ensemble is a.a.s. connected iff $b\rightarrow\infty$
while the survival graph ensemble is a.a.s. disconnected iff $b\rightarrow-\infty$.
\end{theorem}

This theorem provides an explicit dependence of the threshold and asymptotic network breakdown probability
on the node breakdown probability $\epsilon$. {Comparing (\ref{rn2}) with (\ref{rn1}), we can immediately 
see that the effect of the node faults appears in (\ref{rn2}) as a multiplicative factor $\kappa^{-1/2}$. }

\begin{IEEEproof}
Let ${G}(\mathcal{X}_n,g_{r_n})$ denote an instance of a RGGs $\mathcal{G}(\mathcal{X}_n,g_{r_n})$.
We then have
\begin{equation}
{P}_{\Omega_n}(\epsilon;s)\!=\!\mathrm{Pr}[{G}(\mathcal{X}_s,g_{r_n^\ast})\mbox{ is connected}],
\end{equation} 
for any $s\in[s^{(-)}\!,s^{(+)}]$ as $n\!\rightarrow\!\infty$.
Let $\delta_n$ be a sequence that satisfies $\delta_n\ln n\rightarrow 0$ and $\delta_n\sqrt{n}\rightarrow \infty$
as $n\rightarrow\infty$.
From the definition of $s^{(\pm)}$ in Lem.~\ref{lem_typ_bre}, we obtain
\begin{align}
r_n^\ast&=\sqrt{\left(1\pm \frac{\delta_n}{\kappa}\right)\frac{\ln s^{(\pm)}+b-\ln\kappa+o(1)}{s^{(\pm)}}}\nonumber\\
&=\sqrt{\frac{\ln s^{(\pm)}+b-\ln\kappa+o(1)}{s^{(\pm)}}}.  \label{eq_prrg1}
\end{align}
Using Thm.~\ref{thm_con} and monotonicity of connectivity,
 we have $\lim_{n\rightarrow \infty}{P}_{\Omega_n}(\epsilon;s)= 1-e^{-\kappa e^{-b}}$ for any $s\in[s^{(-)},s^{(+)}]$.
By Lem.~\ref{lem_typ_bre} we obtain $\lim_{n\rightarrow \infty}{P}_{\Omega_n}(\epsilon)= 1-e^{-\kappa e^{-b}}$.
Following statements are obtained by taking the $b\rightarrow \pm\infty$ limits.
\end{IEEEproof}

Equivalently, there exists a phase transition threshold of the node breakdown probability $\epsilon$ when $r_n$ is fixed.
\begin{cor}\label{cor_rrg1}
Let us consider the node fault model with node breakdown probability $\epsilon\in [0,1)$.
Assume that $S$ be the unit-area square, 
\begin{equation}
r_n(d)\triangleq \sqrt{\frac{d\ln n}{C n}}\quad (d\ge 1), 
\end{equation}
{with a factor $C$ in~(\ref{eq_rgg_a})}, $\epsilon^\ast= 1/d$, and
functions $g_{r_n}$ and $g$ satisfy conditions (\ref{eq_cond1})-(\ref{eq_cond3}).
Then, for $\mathcal{X}_n$ over $S$ and $\Omega_n=\mathcal{G}(\mathcal{X}_n,g_{r_n(d)})$, we have
$\lim_{n\rightarrow\infty}{P}_{\Omega_n}(\epsilon)\rightarrow 0$ if $\epsilon <\epsilon^\ast$
and $\lim_{n\rightarrow\infty}{P}_{\Omega_n}(\epsilon)\rightarrow 1$ if $\epsilon >\epsilon^\ast$.
\end{cor}

\section{Non-asymptotic analysis}\label{sec_app}

{Although the result of the previous section clearly unveils the asymptotic behavior of the connectivity 
of RGGs with node faults, it cannot be applied directly to a practical assessment in which the number of nodes is finite.
We should be aware of the {\em finite-size effect}, i.e., the discrepancy between infinite and finite graphs.
In this section, we derive a simple approximation formula of the network breakdown probability for
 the general connection model with a finite number of nodes.}

\subsection{Approximation formula for RGGs}
Dettmann and Georgiou recently proposed a general approximation formula of the connection probability for RGGs~\cite{Dett}.
Let us consider a general connection model $\mathcal{G}(\mathcal{P}_\rho,g_{r_\rho})$ defined by the Poisson point process $\mathcal{P}_\rho$
 over the two-dimensional unit-area square.
The approximated value of the connection probability \cite{Dett}, $P_{fc}$,  is given by
\begin{equation} 
P_{fc}\triangleq \exp\left[-\rho e^{-2\pi\rho H_1}- \frac{2}{H_0}e^{-\pi \rho H_1} -\frac{4}{\rho H_0^2}e^{-\pi \rho H_1/2}\right], \label{eq_appr1}
\end{equation}
where $H_m$ ($m=0,1,\dots$) represents the $m$th moment of the connection function which reads
\begin{equation}
H_m\triangleq \int_{0}^{\infty}dr g_{r_\rho}(r) r^m. 
\end{equation}
The approximation formula is based on the fact that the number of isolated nodes asymptotically follows the Poisson distribution.
It contains three terms in the exponent, i.e., the first term corresponds to the bulk (inside of the {domain}), 
the second to the edges,  and the third to the corners.

\subsection{Approximation formula for RGGs with node faults}

Now we consider the node fault model of RGGs.
We introduce additional assumptions to approximate the network breakdown probability for
 the general connection model $\mathcal{G}(\mathcal{X}_n,g_{r_n})$.
The first assumption is that de-Poissonization technique described in Sec.~\ref{con_normal} is applicable for sufficiently large $n$.
We thus apply (\ref{eq_appr1}) to $\mathcal{G}(\mathcal{X}_n,g_{r_n})$ instead of $\mathcal{G}(\mathcal{P}_\rho,g_{r_\rho})$
neglecting the fluctuations of the number of nodes.
As the second assumption, we neglect the fluctuation of the number of nodes in survival graphs in the node fault model, i.e.,
 we replace $\rho$ in (\ref{eq_appr1}) to $\kappa n$, the expectation number of nodes in survival graphs.
This approximation is asymptotically correct as indicated by Lem.~\ref{lem_typ_bre}.
Consequently, the approximation formula of the network breakdown probability in the general connection model $\mathcal{G}(\mathcal{X}_n,g_{r_n})$
is given by
\begin{align}
\tilde{P}_{\Omega_n}(\epsilon)\triangleq& 
1-\exp\left[-\kappa n e^{-2\pi\kappa n H_1}\right.\nonumber\\ 
&\left.-\frac{2}{H_0}e^{-\pi\kappa n H_1} -\frac{4}{\kappa n H_0^2}e^{-\pi \kappa n H_1/2}\right]. \label{eq_appr2}
\end{align}
Neglecting contributions of edges and corners leads to a simplified approximation formula: 
\begin{equation}
\tilde{P}_{\Omega_n}(\epsilon)\simeq 
1-\exp\left[-\kappa n e^{-2\pi\kappa n H_1} \right]. \label{eq_appr3}
\end{equation}

We here validate whether these formulae correctly predict the asymptotic behavior of the network breakdown probability.
Assume that the connection function $g_{r_n}$ satisfies conditions (\ref{eq_cond1})-(\ref{eq_cond3}).
Using change of variables, we have $H_1=Cr_n^2/(2\pi)$ and $H_0=\Theta(r_n)$.
If we set $r_n = \sqrt{(\ln n+b)/(C\kappa n)}$, we find 
\begin{align}
\tilde{P}_{\Omega_n}(\epsilon)=& 
1-\exp\left[-\kappa e^{-b}+O(\ln^{-1/2} n) +O(n^{-1/4}\ln^{-1} n)\right]\nonumber\\
\rightarrow& 1-e^{-\kappa e^{-b}}\quad (n\rightarrow \infty). \label{eq_appr4}
\end{align}
The limiting network breakdown probability agrees with that shown in Thm~\ref{prop_rrg1}.
In addition, we can confirm that the first term in (\ref{eq_appr2}) is dominant.
This fact allows us to use a simple form (\ref{eq_appr3}) for sufficient large $n$.

\section{Rayleigh SISO model}\label{sec_ex}

In this section, we compare the estimation by the approximation formulae,
and the phase transition thresholds to the experimental values obtained by numerical simulations.
{Here, as an example of practical RGG models, we consider 
 single input single output (SISO) wireless communication channels with Rayleigh fading \cite{Coon}.
 This channel model with appropriate parameter well reflects a practical wireless channel.} 

\subsection{Definition and threshold}

We first describe the detail of the model.
To reflect the attenuation in wireless communication channel, we employ a path loss function $r_{ij}^{-\eta}$
where $r_{ij}(>0)$ is the distance between two nodes $i$ and $j$ and $\eta$ represents the path loss exponent.
The fading component is simply modeled as the Rayleigh fading.
In other words, the channel gain $|h|^2$ is assumed to be a random variable that follows an exponential distribution $e^{-|h|^2}$.
If we neglect a shadowing effect and co-channel interference, the signal-to-noise (SNR) ratio is then given by $\beta_0^{-1}r_{ij}^{-\eta}|h|^2$ where
$\beta_0\propto P^{-1}$ is a parameter depending on the transmit power $P$ of a sensor node.
Given that the connection probability between two nodes is defined as
\begin{equation}
g(x) \triangleq \mathrm{Pr}[\mathrm{SNR}\ge \theta], \label{eq_siso1}
\end{equation}
for a constant $\theta(>0)$, 
then the corresponding connection function reads
\begin{equation}
g(x)=e^{-\beta x^\eta}, \label{eq_siso2}
\end{equation}
where $\beta=\theta \beta_0$ is a control parameter.
We call the RGG model with connection function~(\ref{eq_siso2}) the Rayleigh SISO model \cite{Coon}.
The path loss exponent $\eta $ takes $2$ in a free propagation environment without obstacles and around $6$ in a scattered environment~\cite{Erc}.
In addition, the model corresponds to the hard-disk model as $\eta\rightarrow\infty$.

The asymptotic analysis (Sec.~\ref{sec_prob}) and non-asymptotic analysis (Sec.~\ref{sec_app}) are applicable to 
the Rayleigh  model because its connection function satisfies the conditions (\ref{eq_cond1})-(\ref{eq_cond3}).
Before describing the numerical results, we summarize the phase transition phenomena in the model.
The constant $C$ in (\ref{eq_rgg_a}) is given by
\begin{equation}
C_\eta = \frac{2\pi}{\eta}\Gamma\left(\frac{2}{\eta}\right), \label{eq_siso3}
\end{equation}
as a function of $\eta$, where $\Gamma(z)=\int_{0}^\infty t^{z-1}e^{-t}dt$ is a Gamma function.
Then, the phase transition threshold of a parameter $\beta$ with $\epsilon$ and $\eta$ fixed reads
\begin{equation}
\beta^\ast=\left(\frac{C_\eta\kappa n}{\ln n}\right)^{\frac{\eta}{2}}. \label{eq_siso4}
\end{equation}
This threshold shows the relation between the local parameter $\beta$ 
and the asymptotic {connection} probability as a global property.
It also indicates how to control the transmit power $P$ for robust WSNs against {the node faults}.
From the relation $\beta\propto P^{-1}$, we have {$P\propto \kappa^{-\eta/2}$}.
{It suggests that the required transmit power to maintain connectivity is inversely proportional to $(1-\epsilon)^{\eta/2}$.}

\begin{figure}[!t]
\centering
\includegraphics[width=0.95\linewidth]{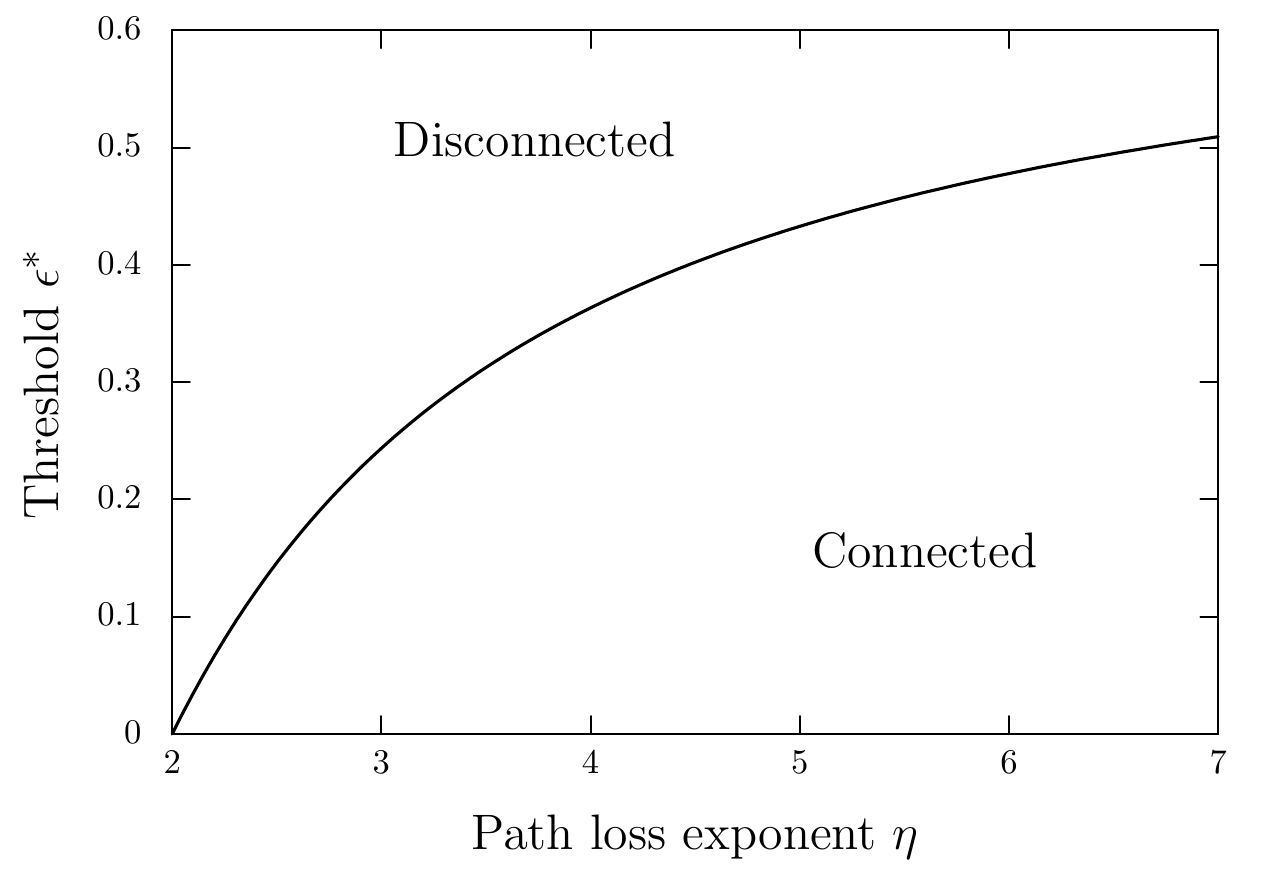}
\caption{Node breakdown probability threshold $\epsilon^\ast$ as a function of the path loss exponent $\eta$
in Rayleigh SISO model when $\beta=\pi(n/\ln n)^{\eta/2}$.
The asymptotic network breakdown probability converges to $1$ if $\epsilon>\epsilon^\ast$
while the survival graphs are a.a.s. connected when $\epsilon<\epsilon^\ast$.
 }
\label{fig_sim0}
\end{figure}

Similarly, the phase transition threshold of the node breakdown probability $\epsilon$
 with $\beta$ and $\eta$ fixed is given by
\begin{equation}
\epsilon^\ast=1-\frac{\ln n}{C_\eta n}\beta^{\frac{2}{\eta}}. \label{eq_siso4a}
\end{equation}
Figure~\ref{fig_sim0} illustrates the relation of the path loss exponent $\eta$ to the threshold $\epsilon^\ast$.
Here we set $\beta=\pi (n/\ln n)^{\eta/2}$,
 which corresponds to the threshold~(\ref{eq_siso4}) with $\epsilon=0$ and $\eta=2$.
We find that the threshold $\epsilon^\ast$ increases as the environment becomes more scattered
when a constant factor of $\beta$ is fixed.

\subsection{Node breakdown probability dependency}

Now we compare our analyses to numerical results.
We first examine the network breakdown probability as a function of the node breakdown probability $\epsilon$.
Here we concentrate on the free-propagation model, i.e., $\eta=2$.

\begin{figure}[!t]
\centering
\includegraphics[width=0.95\linewidth]{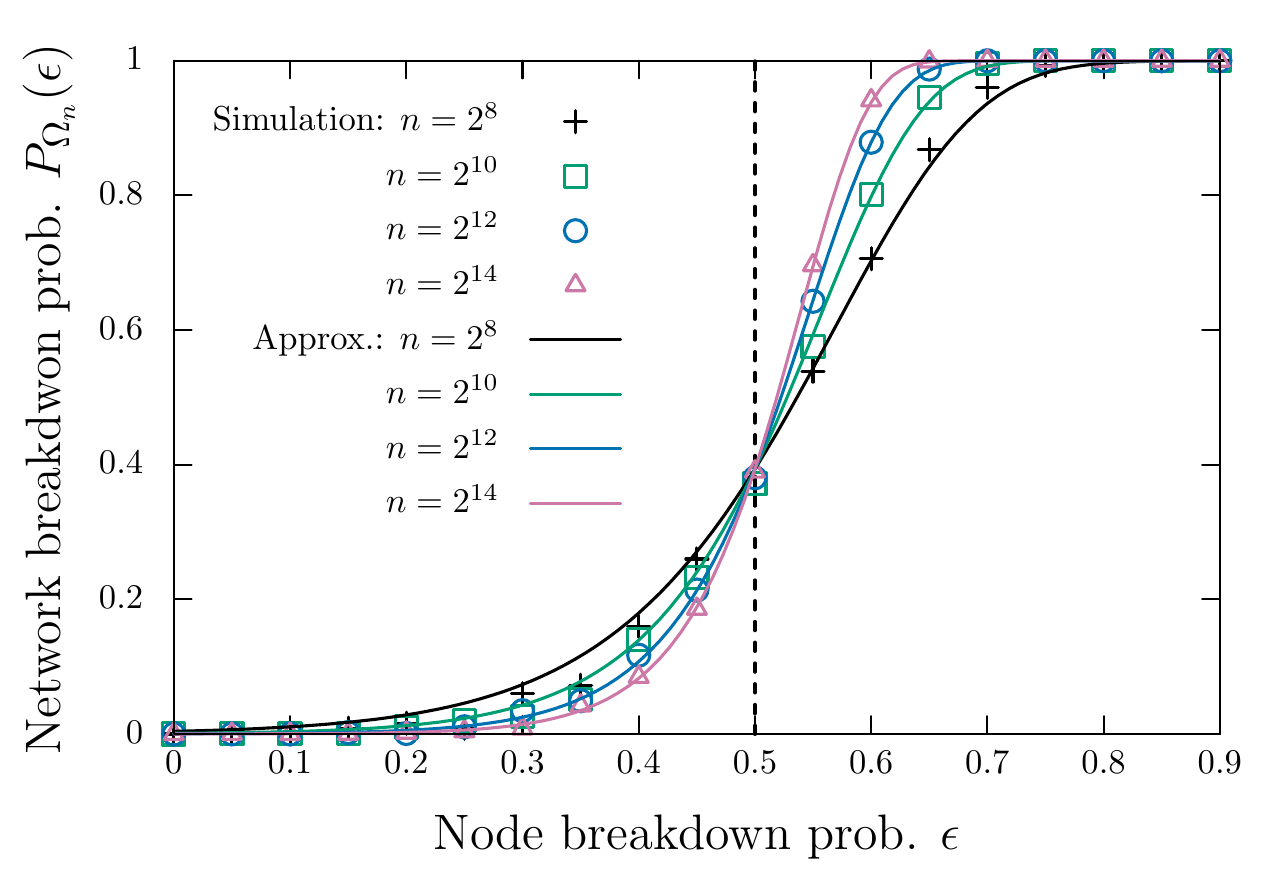}
\caption{Network breakdown probability of Rayleigh SISO model with $\eta=2$ and $\beta=\pi n/(2\ln n)$
 as a function of node breakdown probability $\epsilon$.
 Symbols represent numerical results calculated from $10^4$ survival graphs while
 solid lines correspond to the approximation formulae. The vertical dashed line represents 
 the phase transition threshold $\epsilon^\ast =1/2$.
 }
\label{fig_sim}
\end{figure}

Figure~\ref{fig_sim} shows the network breakdown probability as a function of $\epsilon$ for the Rayleigh SISO model with $\beta=\pi n/(2\ln n)$.
In the numerical simulation, the depth-first search {was} performed to judge whether randomly generated 
survival graphs are connected or not.
The network breakdown probability is then estimated after $10^4$ survival graphs are generated.
To simplify the simulation, we use the toroidal distance instead of Euclidean distance.
In other words, we consider the domain $S$ to be a unit torus.
It results in neglecting the effects of edges and corners of the unit-area square, 
which allows us to use (\ref{eq_appr3}) instead of (\ref{eq_appr2}).
For the Rayleigh SISO model with $\eta=2$, an approximation formula~(\ref{eq_appr3}) reads
\begin{equation}
\tilde{P}_{\Omega_n}(\epsilon)\simeq 
1-e^{-\kappa n^{1-2\kappa}}, \label{eq_appr5}
\end{equation}
which is represented by solid lines in Fig.~\ref{fig_sim}.

The result shows that {approximation values obtained by (\ref{eq_appr5})} well agrees with the numerical results
though some finite-size effects are observed for small $n$.
In addition, the network breakdown probability approaches to the discontinuous phase transition line at $\epsilon^\ast=1/2$ as $n$ grows.

\subsection{Transmit power dependency}

We next investigate the network breakdown probability as a function of a parameter $\beta$ related to the transmit power.
Here we assume that $\eta=4$ and the node breakdown probability $\epsilon$ is fixed.
To rescale the parameter $\beta$, we use the phase transition threshold $\beta_c^\ast$ without {node faults} ($\epsilon=0$).
We thus define a parameter $\delta$ as
\begin{equation}
\beta= \delta \beta_c^\ast,\quad \beta_c^\ast \triangleq \frac{\pi^3n^2}{4\ln^2 n}. \label{eq_siso5}
\end{equation}
Combining (\ref{eq_siso5}) with (\ref{eq_appr3}), {the approximate network breakdown probability reads}
\begin{equation}
\tilde{P}_{\Omega_n}(\delta)\simeq 1-e^{-\kappa n^{1-\frac{\kappa}{\sqrt{\delta}}}}, \label{eq_siso6}
\end{equation}
as a function of the scaling factor $\delta$.

\begin{figure}[!t]
\centering
\includegraphics[width=0.95\linewidth]{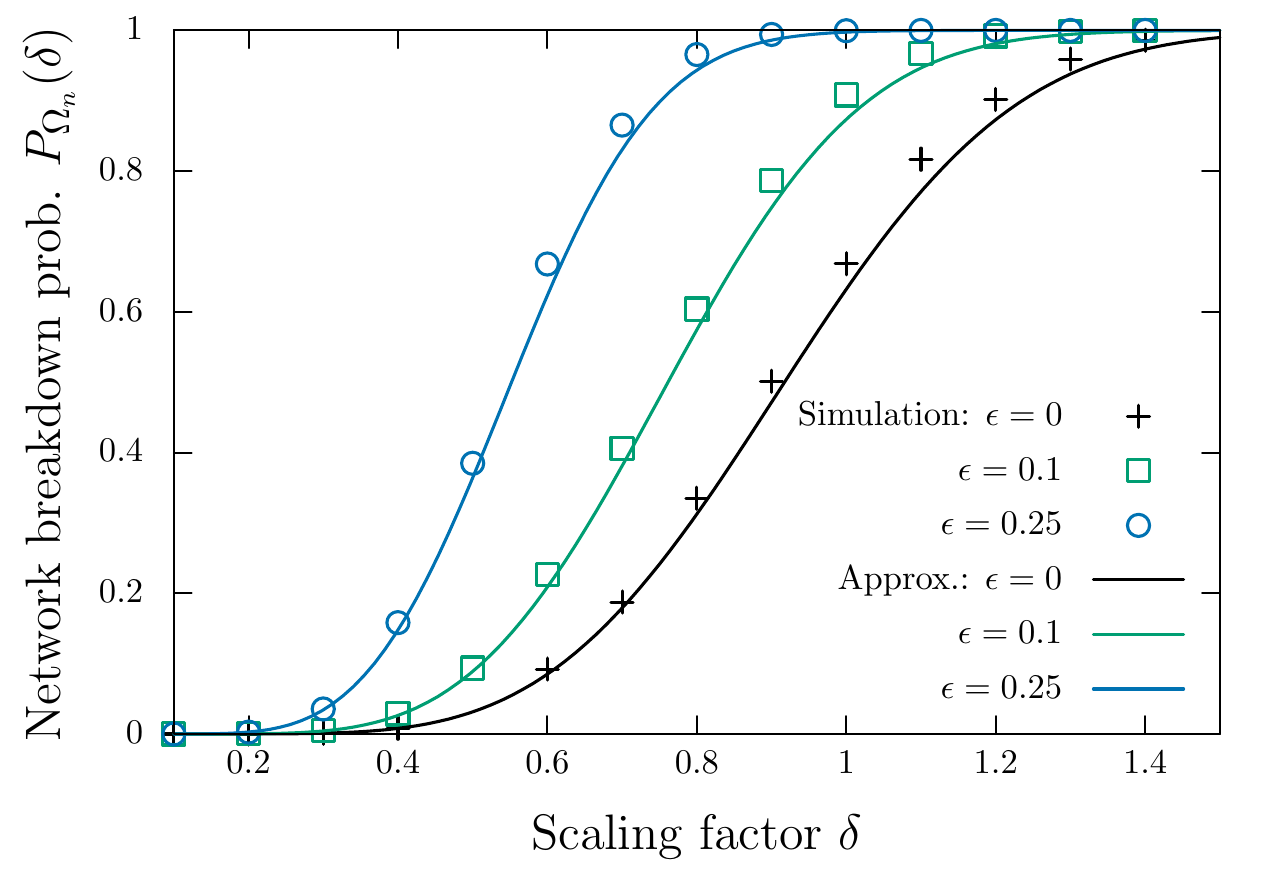}
\caption{Network breakdown probability of Rayleigh SISO model with $\eta=4$, $n=2^{12}$, and $\epsilon=0,0.1,0.25$
 as a function of $\delta =\beta/\beta^\ast_c$.
 Symbols represent numerical results averaged over $4\times 10^4$ survival graphs while
 solid lines correspond to the approximation formulae.
 }
\label{fig_sim1}
\end{figure}

Figure~\ref{fig_sim1} shows the network breakdown probability as a function of $\delta$ with $\epsilon=0,0.1,0.25$.
To examine the practical RGGs rather than an asymptotic situation, we consider finite graphs with $n=2^{12}$.
As with the last subsection, numerical simulations {were} performed using the toroidal distance.
The approximation formula (\ref{eq_siso6}) provides a good estimation of the network breakdown probability.
In particular, {the} finite-size effects of the probability with $\epsilon>0$ is roughly the same as
 that without {node faults}.
It suggests that neglecting fluctuations as additional assumptions to derive (\ref{eq_appr3}) 
has a small contribution to the finite-size effects.
These results indicate that the approximation formulae will be helpful to control the transmit power by evaluating
the network breakdown probability.

\section{Conclusion}

{In this paper, we {studied} the connectivity of random geometric graphs in the node fault model
 as an abstract model of ad hoc WSNs with unreliable nodes.} 
This paper provides two distinct analyses of the model:
(i) the asymptotic analysis for infinite RGGs that reveals phase transition phenomena of connectivity and
its threshold, and
(ii) the non-asymptotic analysis for finite RGGs that provides a useful approximation formula 
of the network breakdown probability.
{We then {compared them to numerical experiments} in the Rayleigh SISO model. 
The asymptotic analysis gives us explicit dependence of the threshold
 on model parameters such as the path loss exponent while 
 the numerical results show that the approximation formulae 
well estimate the network breakdown probability.}

In summary, the {theoretical results obtained in this paper clearly indicate} the relationship between local model parameters and
general connection functions,
and connectivity as a global property of WSNs.
They will be useful analytical tools to design WSNs that is immune to the node faults 
and stimulate further theoretical studies on robust WSNs.


\section*{Acknowledgment}
The authors would like to thank Mr. Wataru Mizuno for executing a part of numerical simulations.
This study {was supported in part} by JSPS KAKENHI Grants Numbers 16K14267, 16H02878 (TW), and 17H06758 (ST).



\end{document}